\documentclass[conference]{IEEEtran}
\IEEEoverridecommandlockouts
\usepackage{endnotes}
\usepackage{color}
\usepackage{graphicx}
\usepackage{wrapfig}
\usepackage{multirow}
\usepackage{listings}
\usepackage{url}
\usepackage{balance}
\usepackage{listings}
\usepackage{verbatim}
\usepackage{subcaption}
\usepackage{stfloats}
\usepackage[numbers]{natbib}
\usepackage{amsmath}
\usepackage{algorithm}
\usepackage{algpseudocode}
\usepackage{booktabs}
\usepackage{listings}
\usepackage{xcolor}
\usepackage{array}
\usepackage{booktabs}
\usepackage{ragged2e}
\usepackage[T1]{fontenc}
\usepackage[htt]{hyphenat}
\hyphenchar\font=`\-
\usepackage[T1]{fontenc}
\usepackage[htt]{hyphenat}
\usepackage{microtype}
\definecolor{mGray}{rgb}{0.5,0.5,0.5}
\definecolor{backgroundColour}{rgb}{0.95,0.95,0.92}
\lstdefinestyle{PythonStyle}{
	language=bash,
	backgroundcolor=\color{backgroundColour},
	commentstyle=\color{mGray},
	keywordstyle=\color{mGray},
	numberstyle=\tiny\color{mGray},
	stringstyle=\color{mGray},
	basicstyle=\ttfamily\linespread{0.9}\small\color{mGray},
	breakatwhitespace=true,
	breaklines=true,
	captionpos=b,
	keepspaces=true,
    numberblanklines=false,
	showspaces=false,
	showstringspaces=false,
	showtabs=false,
	tabsize=2,
    frame=lrtb
}
\def\BibTeX{{\rm B\kern-.05em{\sc i\kern-.025em b}\kern-.08em
    T\kern-.1667em\lower.7ex\hbox{E}\kern-.125emX}}

\newcommand{\name}{Maple}

\newif\ifdraft
\drafttrue
\ifdraft
\newcommand{\zhao}[1]{{\textcolor{cyan}    { ***Zhao:      #1 }}}
\newcommand{\lishan}[1]{{\textcolor{magenta}    { ***Lishan:      #1 }}}
\newcommand{\outline}[1]{{\textcolor{blue}    { ***Outline:      #1 }}}
\newcommand{\revision}[1]{ {\textcolor{red}    {\bf #1 }}}
\else
\newcommand{\zhao}[1]{}
\newcommand{\lishan}[1]{}
\newcommand{\outline}[1]{}
\newcommand{\revision}[1]{}
\fi

\begin{document}

\title{\name{}: A Multi-agent System for Portable Deep Learning across Clusters\thanks{Code and dataset will be released upon publication.}}



\author{
\IEEEauthorblockN{Molang Wu}
\IEEEauthorblockA{
Department of Electrical and Computer Engineering \\
Rutgers University, USA \\
Email: mo.wu@rutgers.edu}
\and
\IEEEauthorblockN{Zhao Zhang}
\IEEEauthorblockA{
Department of Electrical and Computer Engineering \\
Rutgers University, USA \\
Email: zhao.zhang@rutgers.edu}
}

\maketitle

\begin{abstract}
National cyberinfrastructure efforts, such as ACCESS and NAIRR (National Artificial Intelligence Research Resource) Pilot, offer Graphics Processing Unit (GPU) resources across multiple modest-scale clusters to accommodate the increasing need for deep learning (DL) enabled scientific and engineering research. 
Deploying DL workloads across GPU clusters is technically challenging, as users have to compose launch scripts to adapt to the heterogeneous launchers, schedulers, affinity options, DL framework arguments, and environment variables. 
Composing correct launch scripts is error-prone and can easily frustrate users, impeding research or wasting resources.
In this work, we present \name{}, a multi-agent system that generates correct DL scripts with users' natural language input.
\name{} consists of four agents with the functionalities of information extraction, template retrieval, script verification, and error correction.
We evaluate \name{} on nine GPU clusters in the U.S., with five representative deep learning model families, and using four parallel DL training paradigms.
\name{} achieves over 95.6\% accuracy in generating launching scripts across the 567 test cases.
Leveraging multiple language models with an aggregated size of 10~B parameters, \name{} delivers comparable performance to the state-of-the-art models of GPT-5, Claude, and Gemini.
Together, these results highlight \name{}’s practical value in enabling portable and scalable distributed DL across heterogeneous HPC (high-performance computing) environments.

\end{abstract}

\section{Introduction}
The fusion of deep learning and scientific research has made significant breakthroughs across domains, including structural biology~\cite{alphafold3}, materials science~\cite{buzzy2025polymicrosbootstrappingfoundationmodel}, weather forecasting~\cite{graphcast2023}, and many others.
Researchers are also designing AI scientists~\cite{ai_scientist, gottweis2025towards} to automate the literature study, hypothesis formulation, experiment design and execution, results assessment, and report generation. 
The modern DL-enabled science heavily relies on GPU clusters to develop and deploy the models.
To accommodate this new research paradigm, national cyberinfrastructure projects in the U.S., such as ACCESS~\cite{access} and NAIRR Pilot~\cite{NAIRR}, offer GPU resources across several geographically distributed clusters. 
In many cases, a user needs to run the DL workload on multiple GPU clusters, either to avoid the excessively long queue wait time on a cluster~\cite{ding2023mirage} or to leverage the more powerful GPUs on a new machine. 

However, porting DL workloads is extremely challenging, if not prohibitive, across GPU clusters given the heterogeneous hardware/software stack.
Some clusters are equipped with NVIDIA A100 and H100 GPUs, while others have the Grace-Hopper (GH200) superchips or the Intel Max series GPUs.
Even for NVIDIA GPU clusters, the drivers may differ in versions.
These clusters run with different schedulers such as Slurm~\cite{yoo2003slurm} and PBS~\cite{pbs-qsub}.
Although most clusters use lmod~\cite{mclay2011best} to manage software versions, they differ in Python package management tools (i.e., Anaconda and Python Virtual Environment), which are essential for PyTorch~\cite{NEURIPS2019_9015}, the cornerstone software for modern DL research.
The computing centers also operate the clusters with different policies, such as maximum wall clock time.
From the user's perspective, their choice of the launcher also differs between commonly used DL frameworks, such as PyTorch, DeepSpeed~\cite{rasley2020deepspeed}, Megatron-LM~\cite{ShoeybiPatwary2019-megatronlm}, and Accelerate~\cite{accelerate}.
All these vast hardware/software heterogeneities are reflected in the launch script.
To enable portable DL workloads across GPU clusters, a user needs to compose an individual script for each machine in a brute-force way, which is time-consuming.
This challenge impedes the efficient use of GPU resources across clusters for users, especially those with limited HPC experience.
It also prevents the national cyberinfrastructure efforts from offering a uniform user experience across GPU clusters.
Table~\ref{tab:example} shows an abbreviated example of ViT training scripts on Perlmutter and Polaris with a user description to highlight the differences in launch scripts.
\begin{table}[ht]
\centering
\caption{Example launch scrips for ViT training across Perlmutter and Polaris.\label{tab:example}}
\renewcommand{\arraystretch}{1.15}
\setlength{\tabcolsep}{3pt}
\begin{tabular}{
    p{0.25\linewidth}
    >{\RaggedRight\arraybackslash\footnotesize}p{0.35\linewidth}
    >{\RaggedRight\arraybackslash\footnotesize}p{0.35\linewidth}
}
\toprule
\textbf{User Description} & \textbf{Perlmutter} & \textbf{Polaris} \\
\midrule
I want to train ViT using torchrun with 8 GPUs across 2 compute nodes on Perlmutter/Polaris, my training file is run\_image\_classification.py and my training arguments is ...  &
\texttt{srun -N 2 -n 8 bash -c 'torchrun --nnodes=2 --nproc\_per\_node=4 --node\_rank=\$SLURM\_PROCID --master\_addr=\$MASTER\_ADDR --master\_port=29400 run\_image\_classification.py ...'} &
\texttt{mpiexec -n 8 -ppn 4 -hostfile hostfiles.txt -genv MASTER\_ADDR \$MASTER\_ADDR -genv MASTER\_PORT 29500 python -u run\_image\_classification.py ...}
\\
\bottomrule
\end{tabular}
\end{table}

The computing community has long been designing infrastructure to enable efficient portability. 
In Grid Computing, such as TeraGrid~\cite{wilkins2008teragrid} and Open Science Grid~\cite{osg2007}, a user lists the associated commands on each machine in a configuration file. 
Apache Hadoop~\cite{shvachko2010hdfs} and Spark~\cite{zaharia2012resilient} leverage the Java Virtual Machine (JVM) for portability.
Cloud computing providers, such as Amazon Web Services, use virtual machines for consistent software dependencies.
To the same end, recent DL development in the industry heavily uses Kubernetes~\cite{burns2016borg}, Kubeflow~\cite{bisong2019kubeflow}, and Docker~\cite{merkel2014docker}.
The Diamond~\cite{xie2025diamond} service provides the container image and job management functionalities for DL training across GPU clusters. 
Existing solutions address the portability issue from the software environment perspective; none of them target the portable scripts that reflect the hardware/software heterogeneity across GPU clusters.

DL workload launch script generation can be viewed as a code generation problem, which is actively being studied~\cite{roziere2023codellama, li2024starcoder2, hu2024agentcoder}.
We examine the effectiveness of the modern code generation tools (see \S\ref{sec:expr:A}) and find that their capability of generating DL workload launch scripts is limited.
The limited capability is attributed to the lack of cluster-specific knowledge of schedulers, launchers, DL frameworks, node configurations, and management policies.
To address this knowledge gap, we present \name{}, a template-enabled multi-agent system that automatically generates DL workload launch scripts from natural language descriptions. 
At its core, \name{} is enabled by a library of launch script templates that we collect from real DL applications.
\name{} consists of four collaborating agents:
(1) an information extraction agent that interprets user input and identifies key parameters;
(2) a template retrieval agent that searches for the most relevant script templates;
(3) a verification agent that checks the script’s correctness and consistency with system constraints; and
(4) a debugging agent that analyzes and corrects errors if the job fails.
Collectively, these agents form an automated workflow to generate DL workload launch scripts with minimal human effort and ensure portability across heterogeneous clusters.

To evaluate \name{}, we develop a suite of 567 test cases across nine U.S. government-funded clusters, five representative DL model families, and four widely adopted DL frameworks and parallel paradigms.
Our experiments show that \name{} achieves an average accuracy of 95.6\% across all test cases with models of $\sim$10~B parameters.
The \name{} template library also enhances the performance of state-of-the-art LLMs such as GPT-5, Amazon Nova, Claude Sonnet4, and Gemini for DL workload launch scripts generation, with a comparable accuracy to \name{}, though with $\sim$100~B parameters.

The rest of the paper is organized as follows:
We introduce the background of distributed DL workloads on GPU clusters in \S\ref{sec:back}.
We present the system design and implementation details in \S\ref{sec:design} and \S\ref{sec:impl}, respectively.
The experiment results are shown in \S\ref{sec:expr} and we showcase a few common errors in \S\ref{sec:error}.
Related work is discussed in \S\ref{sec:related}.
Finally, we conclude in \S\ref{sec:conc}.

\section{Background}
\label{sec:back}
GPU clusters are critical equipment in modern DL-enabled science and engineering.
Nationwide cyberinfrastructure programs such as ACCESS and NAIRR Pilot offer GPU resources across several computing centers.
Department of Energy (DOE) is also investing in leadership-class computing facilities, e.g., ALCF Aurora and NERSC Perlmutter.
These GPU-dense clusters offer vastly heterogeneous hardware and software.
Users experience difficulties in transition between platforms, and the cyberinfrastructure efforts find it difficult to provide a uniform user experience across clusters. 

The OpenFold~\cite{ahdritz2024openfold} team started the AlphaFold2 reproducibility research on NERSC Perlmutter, which uses SLURM and has four NVIDIA A100 GPUs per node.
Later, they ran out of node hours and had to migrate to TACC Lonestar6 to continue the experiment. 
Lonestar6 also runs SLURM but has three NVIDIA A100 GPUs per node.
A second notable difference is that Perlmutter supports Anaconda, while TACC recommended a native Python build. Thus 
the team had to reconfigure the software dependencies and develop the training script through trial-and-error.
DOE then awarded 200,000 node hours to the project to continue the protein-protein interaction study on ALCF Polaris, which has the PBS scheduler and four NVIDIA A100 GPUs per node.
The team had to adapt the training script to PBS and replace the srun launcher with mpiexec. 

The above example just showcases a few practical obstacles in the portability of DL workloads across modern GPU clusters. 
The heterogeneity spans both hardware and software layers.
On the hardware side, some systems use NVIDIA A100 or GH200 GPUs with high-speed NVLink or InfiniBand interconnects, while others adopt Intel Max series GPUs with different memory hierarchies and communication fabrics like Xe Link. 
Each system connects GPUs in different ways and uses different numbers per node, which changes how DL workloads share data and communicate. 
In addition, every cluster follows its own scheduling and usage policies—for example, job priority rules, queue time limits, and resource allocation constraints—which users must understand before running jobs.
These details are often documented in long user guides, meaning that users have to re-learn the setup process whenever they move to a new machine. 
On the software side, job scheduling and launching environments differ just as widely. 
Some clusters rely on Slurm, others on PBS or custom queueing systems, each with unique syntax and submission form. 
Communication libraries such as MPI, NCCL, or oneCCL also differ in these HPC systems. 
Even the supporting environments like C compiler versions, CUDA toolkits, Python versions, and module systems are different across clusters, often requiring users to rebuild dependencies from the beginning.

As a result, these layers of heterogeneity—spanning hardware connections, system policies, and software environments—create a fragile execution landscape for distributed DL workloads. 
One of the most visible consequences appears in the launch script,  which serves as the interface between users and these complex systems. 
Each launch script depends on users to specify how the job should start, including scheduler options, node counts, GPU mappings, and communication backends. 
And they must be carefully adapted to each system’s unique setup. 
As we observed in our experiments across different clusters, even small differences in schedulers, module environments, or communication libraries can cause scripts that work perfectly on one machine to fail on another. 
These inconsistencies make the launch script a common point of failure in cross-cluster DL workloads, forcing users to repeatedly debug low-level configuration issues instead of focusing on experiments. 
These challenges directly motivate our system design.

In our study, we worked with several major U.S. supercomputing centers, including the Texas Advanced Computing Center (TACC), the National Energy Research Scientific Computing Center (NERSC), the Argonne Leadership Computing Facility (ALCF), the National Center for Supercomputing Applications (NCSA), the Pittsburgh Supercomputing Center (PSC), and Purdue University. 
This wide coverage of production GPU clusters strengthens the generality of \name{}.

\section{System Design}
\label{sec:design}
To enable portable DL workloads across GPU clusters, we present \name{}, a multi-agent system that generates launch scripts with users' natural language inputs.
Conceptually, \name{} provides a conversational interface to users and abstracts the complexity of the hardware/software stack with script templates.
\name{} integrates retrieval-based template synthesis, rule-based verification, and lightweight LLM reasoning to self-correct errors in the launch script and software environment.

\begin{figure*}[t]
    \centering
    \includegraphics[width=\textwidth]{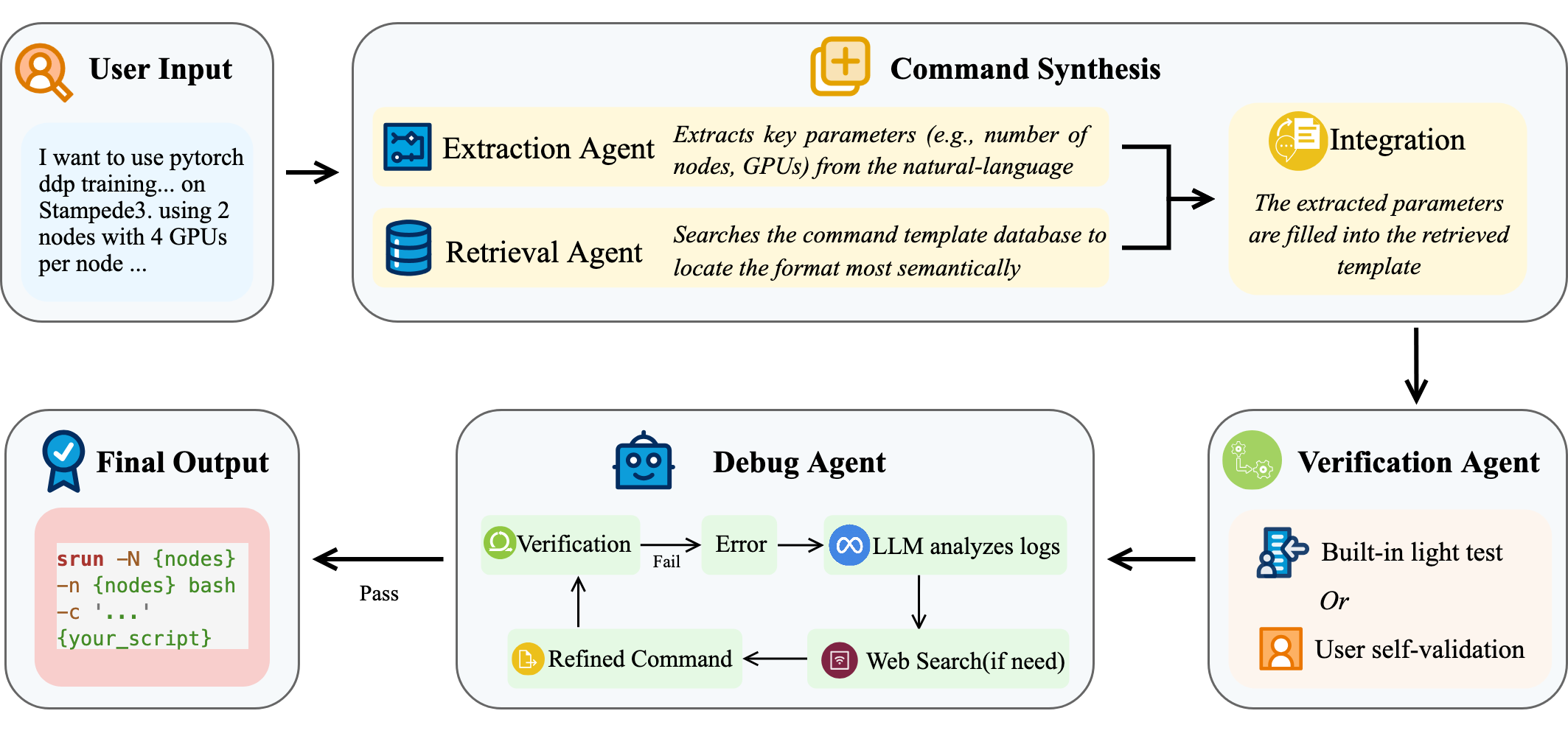}
    \caption{Overview of the system workflow.}
    \label{fig:workflow}
\end{figure*}

Figure~\ref{fig:workflow} shows the overall architecture of \name{}.
\name{} consists of four agents: the extraction agent, the retrieval agent, the verification agent, and the debug agent. 
In a typical generation workflow, a user inputs his/her need in natural language, then the extraction agent extracts critical information from the text. With the identified information, the retrieval agent selects the most relevant templates and produces candidate launch scripts.
After that, the verification agent verifies the correctness of the scripts by running them on the target machine.
Upon success, the generated scripts are returned to users as the final output.
Otherwise, the debug agent leverages the error message from the verification to automatically locate the errors and correct them through trial-and-error.



\subsection{User Input}

The workflow begins with user input, which is the entry point of the system. 
The user input can either be a short description of job needs or a launch script that works on another machine.
In either case, the user input needs to contain sufficient information for \name{} to generate a launch script.
Such information includes the target machine, the DL framework, the parallel strategy, the node count, and the number of GPUs per node.
If the information is not complete within the user's input, \name{} will prompt users to provide missing information.


\subsection{Script Synthesis}
\subsubsection{Extraction Agent}
The extraction agent processes the user input and collects critical information about the target machine, the DL framework, the parallel strategy, and other parameters. 
\name{} leverages a small decoder-only LLM (i.e., SLM, small language models) that is trained to recognize this critical information in plain text. 
There exist many candidate SLMs.
Our selection standard is the comprehensiveness of diverse phrasing styles, such as “two nodes”,  “I’ll use 2 servers”, and "-{}-nodes 2".
The extraction agent also converts both numerical and symbolic expressions into a consistent intermediate format, which ensures the uniform representation for later stages in \name{}.


\subsubsection{Retrieval Agent}
The retrieval agent relies on the extracted information from the extraction agent to fill in the selected script templates.
Specifically, the retrieval agent uses the target machine name to select relevant script templates.
It also selects the DL frameworks (e.g., PyTorch, DeepSpeed, and Accelerate) and the parallel strategy (i.e., data-parallel, fully sharded data parallel, and 3D parallel) by setting the parameters of the scripts.
It uses a lightweight embedding model to semantically match the user’s task description with pre-validated templates collected from real HPC systems. Unlike simple keyword matching, the embedding approach captures contextual similarity. This design enables fast, accurate retrieval even as the repository scales to hundreds of platforms and configuration patterns.
Parameters such as the number of nodes, GPUs per node, master IP address, and communication backend are set by the retrieval agent with extracted information.


\subsubsection{Template Repository}
The template repository is the core of \name{}. 
The heterogeneity in hardware/software across GPU clusters is reflected by the script templates in this repository.
Those templates are collected across nine GPU clusters (see \S\ref{sec:back}) from daily usage. 
Each template represents a verified, runnable script derived from successful real-world tests and matches the practical needs of large-scale DL workloads on GPU clusters. 

We collect 35 templates in total. 
This repository has a good coverage of GPU types, schedulers, DL frameworks, and parallel strategies.
Given the space of this paper, we showcase three example templates. 

\begin{itemize}
\item \textbf{LS6-DDP:} A Slurm-based configuration on Lonestar6 using \texttt{torchrun} for data-parallel (DDP) training:

\begin{lstlisting}[style=PythonStyle, label={lst:ls6}]
srun -N {nodes} -n {nodes} bash -c 'MASTER_ADDR=$(scontrol show hostnames $SLURM_JOB_NODELIST | head -n 1); torchrun --nnodes={nodes} --nproc_per_node={each_node_gpus} --node_rank=$SLURM_PROCID --master_addr=$MASTER_ADDR --master_port={master_port} {your_script}'
\end{lstlisting}

\item \textbf{Aurora-DDP:} A PBS and MPI-based template used on Aurora for distributed data-parallel execution:

\begin{lstlisting}[style=PythonStyle, label={lst:aurora}]
sort -u $PBS_NODEFILE > hostfiles.txt && mpiexec -n {world_size} -ppn {each_node_gpus} -hostfile hostfiles.txt -genv MASTER_ADDR $(head -n 1 hostfiles.txt) -genv MASTER_PORT {master_port} python -u {your_script}
\end{lstlisting}

\item \textbf{DeltaAI-Deepspeed:} A DeepSpeed-based configuration on Delta-AI using Intel MPI for multi-node launch:

\begin{lstlisting}[style=PythonStyle,  label={lst:delta}]
scontrol show hostnames $SLURM_NODELIST | sed 's/$/ slots={each_node_gpus}/' > hostfile.txt && deepspeed --hostfile hostfile.txt --launcher impi {your_script} {deepspeed_config}
\end{lstlisting}
\end{itemize}

The Template Repository serves two key purposes.
First, during generation, it provides verified templates that guide the system in creating runnable scripts matching each cluster’s scheduler and launcher.
Second, during debugging, it acts as a reference, helping the model refine failed scripts by comparing them with similar verified examples.
Over time, newly corrected scripts can be added to the repository, allowing it to improve continuously without retraining.

\subsubsection{Script Synthesis}
Finally, the template retrieved from the repository is combined with the extracted parameters to produce one or more candidate scripts that are ready for verification.

\subsection{Verification Agent}
\name{} proceeds with verification of the generated launch scripts.
Users can certainly verify the generated scripts by trying them out 
in the target GPU cluster.
\name{} provides automated verification via the verification agent using a suite of DL workload mini-apps.
We have developed three mini-apps that combine DL frameworks (i.e., PyTorch, DeepSpeed, and Accelerate) and parallel strategies (i.e., data-parallel, fully sharded data parallel, and 3D parallel). 
These mini-apps are the minimal application codes that can verify the correctness of the parallel DL workload settings.

During automated verification, \name{} starts with the first candidate script and runs the mini-app test. 
If the verification succeeds, \name{} will report results that confirm correctness. 
If the verification fails, the verification agent will capture the standard output and standard error and pass them on to the debug agent for further analysis and refinement.

\subsection{Debug Agent}
When verification fails, the Debug Agent is activated.
It receives the captured error messages and analyzes them using an integrated LLM. 
Here, the debug agent interprets the error trace, identifies likely causes, and produces a refined script suggestion. 
The debugging capability of \name{} is enabled by online retrieval.
The debug agent can retrieve relevant information from multiple data sources, such as a vector database of cluster documentation and online search results in GitHub issues and Stack Overflow. 

The debugging capability is limited by existing knowledge available online and in the documentation.
If the error cannot be resolved in a few iterations, users will need to consult the developers or experts in the computing centers.
However, our results show that \name{} can debug over 90\% of the errors encountered during the launch script development, which evidences the effectiveness of the debugging agent in reducing human effort. 


\subsection{Final Output}
The final stage produces a verified, ready-to-run launch script. 
Once generated, users can directly execute it, save it for reuse, or embed it into Slurm scripts for batch submission.

\section{Implementation}
\label{sec:impl}
In this section, we discuss the implementation details of \name{}.
We will start with model selection for each agent, followed by the template repository.

As introduced in Figure~\ref{fig:workflow}, \name{} consists of four agents: the extraction agent, the retrieval agent, the verification agent, and the debug agent.
We use the Llama-3.2-1B model for the extraction agent, as our measurement shows satisfying accuracy to identify critical information, such as the target machine, launcher, node count, GPU count, and DL framework, from various presentations, including human language and BASH scripts. 
For the retrieval agent, we select all-MiniLM-L6-v2\cite{reimers2019sentencebert}, which has 22 million parameters.
This small language model is efficient and accurate for matching user input with the correct template.
The verification agent does not leverage LLMs.
Instead, it uses a suite of mini-apps to test the correctness of the synthesized or corrected scripts.
The mini-apps are developed based on the GPT-2 training applications with DDP, FSDP, ZeRO-3.
These mini-apps are with trivial computation and one round of allreduce and one round of allgather communication.
They are sufficient to test the correctness of the generated scripts for distributed DL workloads.
The debug agent is enabled with the Llama-3-8B-Instruct model to analyze error logs, suggest corrections, and generate new scripts for testing.
The debug agent can retrieve online information to search for up-to-date troubleshooting information on GitHub issues or Stack Overflow.
This capability allows the system to diagnose emerging framework or hardware issues that are not yet captured in the local knowledge base. 
All the default language models are selected based on the needs of \name{}. 
They can be replaced with larger models, such as GPT-5 and Gemini, through API calls.


The template repository is implemented as a JSON file, which is easy to extend.
It contains templates for popular frameworks such as PyTorch, DeepSpeed, and Accelerate, as well as the main parallelization strategies like DDP, FSDP\cite{zhao2023fsdp}, and ZeRO\cite{rajbhandari2020zero}. Because all templates were validated on actual systems, the database reflects practical usage instead of synthetic examples.

\section{Experiments}
\label{sec:expr}
We performed three experiments to evaluate the design effectiveness of \name{}.
In the first experiment, we test \name{}'s capability of launching script generation across multiple GPU clusters with common parallel DL strategies.
The second experiment evaluates the portability across DL applications and the debug capability of \name{}.
Finally, we compare the debugging capability between \name{} and other open LLMs to study the effectiveness boundary.

\subsection{Cross-Platform and Parallelization Verification}
The first experiment is to examine how \name{} works across GPU clusters and parallel DL strategies.
We use GPT-2~\cite{gpt2} as the benchmark model in all cases, as it is widely supported across major deep learning frameworks, lightweight enough for consistent deployment on heterogeneous clusters, and representative of modern transformer-based architectures used in large-scale distributed DL workloads.
We evaluate \name{} across nine clusters using four launch strategies: PyTorch DDP, PyTorch FSDP, DeepSpeed ZeRO-3, and Accelerate DDP.
As shown in Table~\ref{tab:HPC}, almost all cases passed successfully, with only three exceptions. 
On Aurora, Accelerate DDP failed due to the conflict between the PBS scheduler and the launcher.
On Vista and DeltaAI, ZeRO-3 could not start because the compilation of NVIDIA Apex failed on GH200s. 
A further analysis on the compilation error is in \S\ref{sec:error:B}.

\label{sec:expr:A}
\begin{table}[htbp]
\centering
\caption{Cross-platform verification of parallelization strategies.}
\label{tab:HPC}
\begin{tabular}{lcccc}
\hline
\textbf{Platform} & \textbf{DDP} & \textbf{FSDP} & \textbf{ZeRO-3} & \textbf{Acc-DDP} \\
\hline
Lonestar6   & $\circ$ & $\circ$ & $\circ$ & $\circ$ \\
Perlmutter  & $\circ$  & $\circ$ & $\circ$ & $\circ$ \\
Stampede3   & $\circ$  & $\circ$ & $\circ$ & $\circ$ \\
Vista       & $\circ$  & $\circ$ & $\times$ & $\circ$ \\
Aurora      & $\circ$  & $\circ$ & $\circ$ & $\times$ \\
Delta       & $\circ$  & $\circ$ & $\circ$ & $\circ$ \\
DeltaAI     & $\circ$  & $\circ$ & $\times$ & $\circ$ \\
Anvil       & $\circ$  & $\circ$ & $\circ$ & $\circ$ \\
Bridges-2   & $\circ$  & $\circ$ & $\circ$ & $\circ$ \\
\hline
\end{tabular}
\end{table}

To isolate the source of the effectiveness and to compare our model choice of the extraction and retrieval agents, we compare \name{} against open LLMs with and without template retrieval using the same test suite.
The selected baselines cover a wide range of model families and sizes. 
They include smaller models such as GPT-4 Mini, mid-scale models like Claude Sonnet-4, DeepSeek-R1-70B\cite{deepseekr1}, Qwen-2.5-VL-72B\cite{qwen2023}, and GPT-5, Amazon Nova, Gemini. 
The tests include models with chain-of-thought reasoning, models that give direct outputs, and some trained with extra code data. 
This mix reflects the current landscape of LLMs: some optimized for efficiency, some specialized for code, and others trained with massive datasets to achieve state-of-the-art performance. 
By testing across these models, we ensure that our comparison represents the mainstream choices that users are most likely to try in practice, and highlights the portability advantage of our system across heterogeneous HPC settings. 
Table~\ref{tab:llm_without_retrieval} summarizes the overall success rate of the baselines without template retrieval.
\name{} correctly generates 33 out of 36 cases with a success rate of 91.7\%.
In contrast, GPT-5 has only four success cases (11.1\%), while GPT-4-mini fails in all cases (0\%). 
Qwen-2.5-VL-72B, DeepSeek-R1-70B, Amazon Nova, and Gemini each succeed in only two cases (5.6\%). 
Claude Sonnet-4 achieves the highest success rate of 22.2\% among the open models.


\begin{table}[ht]
\centering
\caption{Accuracy of different models on nine HPC machines across four parallelism paradigms (without retrieval).}
\label{tab:llm_without_retrieval}
\begin{tabular}{lcccc}
\toprule
\textbf{Model} & \textbf{\#Tests} & \textbf{\#Success} & \textbf{Fail} & \textbf{Accuracy} \\
\midrule
GPT-5          & 36 & 4  & 32 & 11.1\% \\
GPT-4 Mini     & 36 & 0  & 36 & 0.0\%  \\
Qwen 2.5 VL 72B & 36 & 2 & 34 & 5.6\%  \\
DeepSeek-R1 70B & 36 & 2 & 34 & 5.6\%  \\
Amazon Nova    & 36 & 2  & 34 & 5.6\%  \\
Claude Sonnet4 & 36 & 8  & 28 & 22.2\% \\
Gemini         & 36 & 2  & 34 & 5.6\%  \\
\midrule
\textbf{Ours}  & 36 & 33 & 3  & \textbf{92.0\%} \\
\bottomrule
\end{tabular}
\end{table}

\begin{table}[ht]
\centering
\caption{Accuracy of different models on nine HPC machines across four parallelism paradigms (with retrieval).}
\label{tab:llm_comparison_retrieval}
\begin{tabular}{lcccc}
\toprule
\textbf{Model} & \textbf{\#Tests} & \textbf{\#Success} & \textbf{Fail} & \textbf{Accuracy} \\
\midrule
GPT-5          & 36 & 33 & 3  & 92\% \\
GPT-4 Mini     & 36 & 16 & 20 & 44\% \\
Qwen 2.5 VL 72B & 36 & 24 & 12 & 67\% \\
DeepSeek-R1 70B & 36 & 24 & 12 & 67\% \\
Amazon Nova    & 36 & 33 & 3  & 92\% \\
Claude Sonnet4 & 36 & 33 & 3  & 92\% \\
Gemini         & 36 & 33 & 3  & 92\% \\
\midrule
\textbf{Ours}  & 36 & 33 & 3  & \textbf{92\%} \\
\bottomrule
\end{tabular}
\end{table}

The key difference between \name{} and the baselines is the script template retrieval.
To isolate the effectiveness of the templates, we run another set of experiments with script template retrieval for open LLMs. 
For very large models that cannot easily connect to an external database, we place the script template directly into the prompt as a simple form of retrieval. 
The results are shown in Table~\ref{tab:llm_comparison_retrieval}. With this setup, open models such as GPT-5, Claude, Amazon Nova, and Gemini are able to reach 91.7\% accuracy, the same as \name{}. 
The success rates of DeepSeek-R1-70B and Qwen-2.5-VL-72B also improve from 5.6\% to 67\%, but still fail in 12 cases because the generated scripts miss key parts of the template. 
GPT-4-mini succeeds in 16 out of 36 cases. 
This experiment further confirms that it is the script template retrieval that contributes to the high success rate of \name{} and other baseline LLMs.
Comparing with open LLMs, \name{} achieves the highest accuracy with a much smaller model size.

\subsection{Portability and Debugging Verification}
In the second experiment, we focus on testing portability and debugging capability. 
Unlike fixed benchmarks, real HPC jobs often involve models and tasks that the system has not seen before. 
We select five representative architectures across different categories of deep learning workloads: 
GPT-2 XL is a large language model with about 1.5 billion parameters, designed for text generation tasks. 
The second model is meta-llama/Llama-3.1-8B, which is larger and more advanced than GPT-2 XL, serving as a representative of modern large-scale training. 
BERT-base (Bidirectional Encoder Representations from Transformers) is the third model and is an encoder model widely used for natural language understanding tasks such as text classification. 
On the vision side, we choose ViT-base (Vision Transformer), which demonstrates transformer-based methods for image recognition, and ResNet-18, a convolutional neural network, representing classical CNN-style image classification.
These tests also cover multiple parallel training methods, including DDP and FSDP. 
More importantly, they cannot be solved by simply filling in templates. 
In many cases, the generated scripts have to be further refined, e.g., adjusting the order of options, adding missing arguments such as batch size or tokenizer path, handling dataset paths, or fixing preprocessing steps. 
Similar to the first experiment, we study the effectiveness of \name{} with and without template retrieval in \S\ref{sec:res:woret} and \S\ref{sec:res:wret}, respectively.
This experiment ensures that the evaluation reflects realistic HPC usage, where debugging and adaptation are necessary beyond template completion.

\begin{table}[ht]
\centering
\caption{Script generation accuracy without retrieval.}
\label{tab:models_no1}
\begin{tabular}{lccc}
\toprule
\textbf{Model} & \textbf{Tests} & \textbf{Success} & \textbf{Accuracy} \\
\midrule
GPT-5          & 45 & 13 & 28.9\% \\
GPT-4 Mini     & 45 & 0  & 0.0\%  \\
Qwen 2.5 VL 72B & 45 & 6  & 13.3\% \\
DeepSeek-R1 70B & 45 & 6  & 13.3\% \\
Amazon Nova    & 45 & 11 & 24.4\% \\
Claude Sonnet4 & 45 & 15 & 33.3\% \\
Gemini         & 45 & 13 & 28.9\% \\
\midrule
\textbf{Ours}  & 45 & 43 & \textbf{95.6\%} \\
\bottomrule
\end{tabular}
\end{table}

\begin{table}[ht]
\centering
\caption{Script generation accuracy in five corrected without retrieval.}
\label{tab:models_no2}
\begin{tabular}{lccc}
\toprule
\textbf{Model} & \textbf{Pass@1} & \textbf{Pass@5} & \textbf{Fail} \\
\midrule
GPT-5           & 8 (17.8\%)  & 5 (11.1\%)  & 32 (71.1\%) \\
GPT-4 Mini      & 0 (0.0\%)   & 0 (0.0\%)   & 45 (100\%) \\
Qwen 2.5 VL 72B & 4 (8.9\%)   & 2 (4.4\%)   & 39 (86.7\%) \\
DeepSeek-R1 70B & 4 (8.9\%)   & 2 (4.4\%)   & 39 (86.7\%) \\
Amazon Nova     & 4 (8.9\%)   & 7 (15.6\%)  & 34 (75.6\%) \\
Claude Sonnet4  & 16 (35.6\%) & 0 (0.0\%)   & 29 (64.4\%) \\
Gemini          & 4 (8.9\%)   & 9 (20.0\%)  & 32 (71.1\%) \\
\midrule
\textbf{Ours}   & \textbf{33 (73.3\%)} & \textbf{10 (22.2\%)} & \textbf{2 (4.4\%)} \\
\bottomrule
\end{tabular}
\end{table}

\subsubsection{Without Retrieval}
\label{sec:res:woret}
Table~\ref{tab:models_no1} and Table~\ref{tab:models_no2} present the results of launching script generation accuracy without retrieval. 
For the failures, we allow every model up to five iterations of debugging, where the error message is fed back to the debug agent to produce new launch scripts, then try again.

Without script template retrieval, most mainstream LLMs show very low success rates.
In particular, GPT-4 Mini fails all cases.
GPT-5 and Gemini show achieve 28.9\% accuracy.
Claude Sonnet-4 shows the highest accuracy of 33.3\%.
But many of the generated scripts are unnecessarily complicated, which requires users to follow a series of long instructions instead of a concise script. 
In contrast, \name{} succeeds in 33 test cases for its first generation.
With five debugging iterations using the debug agent, it correctly produces scripts for another 10 cases, which aggregates a 95.6\% accuracy.

Originally, we expected the LLM in the debug agent to look into the error traces and find the root cause, then produce corrected scripts.
However, with our analysis of the erroneous scripts generated by the baseline models, we find that most errors are in the launching script.
Typical errors include incorrect argument setting, communication backend selection, and wrong scheduler options.
This observation indicates that the LLM is struggling to generate correct parameters for the launching script.
Even though the LLM may be able to find the root cause, the debugging capability is hindered by improper parameter generation.
This observation further justifies the script template approach of \name{}, which enforces the parameter structure in launching scripts and achieves higher accuracy.

The 10 cases that \name{} solves with the debug agent show errors such as mismatched CUDA version, incompatible GCC compiler, incorrect communication libraries, missing modules, and failing to propagate environment variables across nodes.
There are also external configuration issues, such as HuggingFace authentication, dataset permissions, and a wrong configuration file path.
These errors are not directly related to GPU clusters, but \name{} still needs to resolve them for users to run successfully.
Finally, a few errors are DL model-specific. 
E.g., ResNet requires proper image resizing and normalization parameters, while BERT for GLUE tasks needs the correct tokenization step.
In all aforementioned errors, the \name{} debug agent suggests clear fixes, e.g., correcting the software version, adding module load commands, exporting environment variables, or applying the right preprocessing routines.
There are two failures that cannot be fixed by the debug agent.
Both failures are on the Intel GPU clusters, Stampede3 and Aurora.
In both cases, the problem stems from the outdated PyTorch training scripts in the official repository, which lack support for Intel’s communication libraries. 
So even after five debugging iterations, the generated launching scripts still fail.

\subsubsection{With Retrieval}
\label{sec:res:wret}
With retrieval, we run the same set of experiments as in \S\ref{sec:res:woret} across \name{} and the baseline LLMs. 
The results are shown in Table~\ref{tab:models_reterival} and Table~\ref{tab:model_reterival2}.

\begin{table}[ht]
\centering
\caption{Script Generation Accuracy with Retrieval.}
\label{tab:models_reterival}
\begin{tabular}{lccc}
\toprule
\textbf{Model} & \textbf{\#Tests} & \textbf{\#Success} & \textbf{Accuracy} \\
\midrule
GPT-5          & 45 & 43 & 95.6\% \\
GPT-4 Mini     & 45 & 10 & 22.2\% \\
Qwen 2.5 VL 72B & 45 & 24 & 53.3\% \\
DeepSeek-R1 70B & 45 & 22 & 48.9\% \\
Amazon Nova    & 45 & 43 & 95.6\% \\
Claude Sonnet4 & 45 & 43 & 95.6\% \\
Gemini         & 45 & 43 & 95.6\% \\
\midrule
\textbf{Ours}  & 45 & 43 & \textbf{95.6\%} \\
\bottomrule
\end{tabular}
\end{table}

\begin{table}[ht]
\centering
\caption{Script Generation Accuracy with Debugging Iterations with Retrieval.}
\label{tab:model_reterival2}
\begin{tabular*}{\linewidth}{@{\extracolsep{\fill}}lccc}
\toprule
\textbf{Model} & \textbf{Pass@1} & \textbf{Pass@5} & \textbf{Fail} \\
\midrule
GPT-5          & 33 (73.3\%) & 10 (22.2\%) & 2 (4.4\%) \\
GPT-4 Mini     & 6  (13.3\%) & 4  (8.9\%)  & 35 (77.8\%) \\
Qwen 2.5 VL 72B& 6  (13.3\%) & 18 (40.0\%) & 21 (46.7\%) \\
DeepSeek-R1 70B& 4  (8.9\%)  & 18 (40.0\%) & 23 (51.1\%) \\
Amazon Nova    & 33 (73.3\%) & 10 (22.2\%) & 2 (4.4\%) \\
Claude Sonnet4 & 33 (73.3\%) & 10 (22.2\%) & 2 (4.4\%) \\
Gemini         & 33 (73.3\%) & 10 (22.2\%) & 2 (4.4\%) \\
\midrule
\textbf{Ours}  & \textbf{33 (73.3\%)} & \textbf{10 (22.2\%)} & \textbf{2 (4.4\%)} \\
\bottomrule
\end{tabular*}
\end{table}

Large baseline models like GPT-5, Nova, Claude, and Gemini show higher accuracies (95.6\%) than mid-sized models of Qwen 2.5 VL 72B and DeepSeek-R1 70B.
However, these large baseline models often come with clear drawbacks: 
Although they take fewer or equal iterations to generate correct scripts compared to \name{}, the response time is much longer (see Table~\ref{tab:model_error_stats}), making the interface less responsive. 
In addition, their generated scripts are often unnecessarily verbose: 
The baseline models rewrite large parts of the script or add unnecessary operations, which makes the workflow more complex and harder to understand. 
These models also tend to produce long explanations alongside the scripts. Such explanations add overhead and sometimes distract from the simple task of script generation. 
In contrast, \name{} achieves the same accuracy with shorter response time, concise outputs, and scripts that are structured and ready to use.

\begin{table}[ht]
\centering
\caption{Comparison of Average Error and Variance across Models.}
\label{tab:model_error_stats}
\begin{tabular}{lcc}
\toprule
\textbf{Model} & \textbf{Mean (seconds)} & \textbf{Stdev} \\
\midrule
GPT-5          & 24.0 & 3.2 \\
Claude         & 13.5 & 3.0 \\
Nova           & 11.6 & 3.2 \\
Gemini         & 9.1  & 2.5 \\
\midrule
\textbf{Ours}  & \textbf{4.7} & \textbf{1.6} \\
\bottomrule
\end{tabular}
\end{table}

Compared the baseline models among themselves, we see that mid-sized models such as Qwen-2.5-VL-72B and DeepSeek-R1-70B show higher accuracies compared to the retrieval-free cases, though their accuracies still stay around 50\%. 
The main reason is that they do not show effectiveness in the debugging stage, often repeating explanations but not giving clear or correct fixes. 
This observation indicates that many errors can not be fully resolved solely with LLMs. 
Both Qwen and DeepSeek show long response times, as they apply chain-of-thought reasoning to explain their steps.
However, the suggestions are often impractical, and the longer wait does not lead to better generation.

GPT-4 Mini shows the weakest results. 
Its size was too small to reliably generate correct scripts, and even when retrieval helps produce partial outputs, it struggles to debug them. 
In most cases, it repeats the same incomplete script rather than correcting the error.

In summary, structured and accurate launch scripts are essential because even small errors in the environment or settings can prevent DL workloads from starting.
With the aid of the debug agent, \name{} shows a high accuracy of 95.6\%.
\name{} can automatically fix most errors that appear in practice. 
The experiments also show that \name{} is highly portable: it works across nine GPU clusters and supports a wide range of DL frameworks. 
This makes distributed DL workloads efficient and accessible to users with all levels of HPC expertise.
 
\subsection{Model Size Comparison}
\begin{table}[htbp]
\centering
\caption{Model size comparison.}
\label{tab:model_size}
\begin{tabular}{lc}
\hline
\textbf{Model/System} & \textbf{Approx. Size} \\
\hline
MiniLM (retrieval)        & 22M      \\
LLaMA-3.2-1B (extraction) & 1B       \\
LLaMA-3.1-8B-instruct (debugging)    & 8B       \\
\textbf{Our system (total)} & \textbf{$<$10B} \\
\hline
GPT-4-mini           & 8B+      \\
DeepSeek-R1-70B      & 70B      \\
Qwen-2.5-VL-72B      & 72B      \\
GPT-5                & 100B+    \\
Gemini       & 100B+    \\
Nova         & 100B+    \\
Claude Sonnet-4         & 100B+    \\
\hline
\end{tabular}
\end{table}

Table~\ref{tab:model_size} compares the size of \name{} with several widely used large language models. 
\name{} is not a single large model but a pipeline of three small ones: MiniLM for template retrieval, LLaMA-3.2-1B for parameter extraction, and LLaMA-3.1-8B-instruct for debugging. Together, the total size is less than 10B parameters.
In contrast, general-purpose models are much larger. GPT-5, Gemini, and Nova are all over 100B parameters. 
Claude Sonnet-4, DeepSeek-R1-70B, and Qwen-2.5-VL-72B are around 70B. 
Despite the large model size, these models still require the script templates retrieval to produce correct launch scripts.

The smaller size of our system brings advantages. 
\name{} can run directly on HPC clusters without external calls, so users can get answers immediately without using Internet connections. 
Because everything runs locally, user data and job information remain private and secure. 
MCP (model context protocol) connections can optionally use online tools when needed, but the core workflow works locally and remains efficient. 
The lightweight design also reduces cost and makes deployment much easier.

\section{Error Analysis}
\label{sec:error}
While \name{} greatly improves the portability and automation of distributed DL workloads, several errors still occur during real-world testing. 
These issues show that even with \name{}, running large-scale jobs across different HPC systems remains challenging. 
To better understand these challenges, we analyze all failed and partially successful runs. 
The goal of this analysis is to identify limiting factors of portability and the problems that novice users are most likely to encounter in practice. 
Based on our findings, we group the errors into three main categories: (1) HPC  environment-related issues; (2) framework-related issues; and (3) user-related issues. 
This section presents typical examples from each category and explains how \name{}'s retrieval, verification, and debug agents help locate and fix these problems.

\subsection{HPC environment-related issues}
Environment-related issues are among the most common sources of failure in distributed DL workloads. 
Most of these problems arise from inconsistencies in communication libraries, driver versions, or system environments that differ between GPU clusters.

One representative error on the DeltaAI cluster is that multi-node training consistently fails because environment variables are not properly propagated to all nodes in the allocation. 
Although the first node is configured correctly, other nodes can not locate the required Python paths, CUDA toolkits, or custom libraries. 
As a result, each node launches with incomplete configurations, leading to immediate runtime errors and job termination.
At first, this problem is difficult to diagnose because the error logs only report missing modules without revealing the root cause. 
The system’s debug agent quickly suggests adding explicit export statements (e.g., PYTHONPATH, LD\_LIBRARY\_PATH) within the srun command to ensure environment consistency across nodes. 
However, based on prior experience on other GPU clusters where SLURM automatically propagated environment variables, we initially dismissed this explanation. 
After repeated failures, we confirm that DeltaAI uses a customized SLURM configuration that does not automatically export user-defined variables between nodes. 
Once we apply the refined script produced by the debug agent, the training executes successfully on all nodes.

Similar problems appear on other clusters. 
On Stampede3, a mismatch between the Intel driver and communication library causes inter-node failures.
On Aurora, version conflicts between compilers and SYCL libraries break initialization. 
These examples highlight that even with correct launch scripts, system-level inconsistencies remain major obstacles to portability.

\subsection{Framework-related issues}
\label{sec:error:B}
Framework-level problems often arise from incompatibilities between deep learning libraries and specific hardware or communication backends. 
One representative case is Vista and DeltaAI, which are equipped with NVIDIA GH200 GPUs.
Distributed training repeatedly fails during the compilation of NVIDIA Apex. At first, the debug agent suspects a configuration error and tries several local fixes, but all attempts fail. 

After several failed attempts, the debug agent invokes online retrieval to access the most up-to-date information. 
Through this connection, the model accesses up-to-date information about hardware compatibility and identifies that the issue stems from a mismatch between the installed PyTorch version and the CUDA toolkit on GH200 systems.
It then recommends switching to the latest nightly PyTorch build. 
Once the environment is rebuilt using this version, the training jobs compile and execute successfully across all nodes.
Similar framework-level incompatibilities are also observed on Intel Max GPUs, where unsupported communication backends cause initialization errors. 
These errors are resolved after guided version adjustments from the debug agent.

\subsection{User-related issues}
User-related issues mostly come from missing configurations or incompatible software settings.
A common example is the mismatch between GCC and CUDA versions on Perlmutter, which causes compilation or runtime errors during setup. 
After examining the logs, the debug agent identifies the incompatibility and suggests the correct environment modules to load (e.g., module load gcc/13.2.0 cuda/12.4), allowing users to fix the issue manually without deep system knowledge. 

Another case involves a missing dataset path or training arguments in the launch script on LS6. 
In one instance, we forget to specify the dataset path and the task name when training BERT, which causes immediate termination. 
After we provided the dataset name and path, the debug agent automatically refined the script with the correct arguments and relaunch the job successfully.

Other user-side mistakes, such as missing authentication for private model downloads or incorrect script paths, occur less often but follow similar patterns.  
In each case, the \name{} debug agent quickly identifies the issue and provides a correct suggestion, significantly reducing manual troubleshooting effort.
More importantly, these capabilities allow non-expert users to resolve complex errors independently, improving both accessibility and usability of the GPU clusters.

\subsection{Unresolved Cases}
Despite these successes, a few unresolved cases remain.
On the NVIDIA GH200 cluster, such as Delta-AI, DeepSpeed Zero-3 cannot be executed because the latest streaming processor architectures were not yet supported in the official release.
It prevents the compilation of fused-adam and other CUDA kernels. 
Similarly, PyTorch’s native DDP was incompatible with Intel XPU devices due to missing upstream support for the Intel backend. 
These issues were beyond the scope of \name{}, as they are from limitations in the underlying frameworks or specific drivers rather than configuration errors. 
They highlight that some portability barriers can only be addressed through official updates, not just automation or debugging improvements. 

In summary, this error analysis highlights that despite substantial progress in automation, distributed DL workloads across heterogeneous GPU clusters remain sensitive to both environmental variation and user configuration, emphasizing the need for adaptive and intelligent tools to bridge this gap.

\section{Related Work}
\label{sec:related}
As deep learning models continue to be adopted for science and engineering research, distributed DL workloads have become essential for large-scale research. 
However, launching jobs across heterogeneous HPC systems remains challenging because their scheduling and runtime environments differ widely. This challenge is a long standing problem of software portability in distributed computing. 

Early distributed computing often relies on manual scripting, where users  specify resources, schedulers, and communication backends directly. 
As systems grow more complex, researchers attempt to simplify this process through standardized interfaces such as GRAM and DRMAA\cite{rajic2007drmaa}, while tools like Condor\cite{thain2005condor} redirect environment-specific operations to improve compatibility. 
The later emergence of cloud computing\cite{armbrust2010view} and virtualization\cite{rosenblum2005virtual} further advances portability by providing isolated environments and standardized APIs, allowing workloads to move more flexibly between data centers. 
Building on these developments, containerization frameworks such as Docker \cite{merkel2014docker} and Singularity \cite{kurtzer2017singularity} extend these ideas to scientific and HPC contexts, enabling consistent runtime environments across heterogeneous systems. 
In parallel, data management infrastructures like Globus \cite{foster2011globus} enable data transfer and job submission across supercomputers. 
However, the differences in hypervisors, file systems, and resource management policies still prevent full portability between cloud and HPC platforms. 
Although virtualization is useful for isolating environments in cloud computing, it remains restricted or discouraged on many HPC systems due to security concerns, administrative policies, and potential performance overhead. 
As a result, they do not fully resolve the portability challenges faced by large-scale distributed DL workloads on HPC clusters.

With the rapid rise of deep learning, these long-standing portability challenges have resurfaced at a much larger scale. 
Training modern foundation models requires coordinated use of hundreds to even hundreds of thousands of GPUs across clusters, and the need to adapt jobs to different HPC environments has made job launching more complex than ever. Frameworks such as PyTorch introduced the distributed launcher torchrun, which automates many of the low-level steps in launching DL jobs: setting up rendezvous, initializing communication backends, spawning processes, and exporting environment variables automatically~\cite{li2020pytorchddp}. Building on this foundation, DeepSpeed extended the idea by introducing a template-based launcher that fills in arguments according to user configurations.
It automatically constructs scripts for large-scale jobs, integrating optimizer states, ZeRO partitioning\cite{rajbhandari2020zero}, and model sharding settings into the launch phase. In contrast, Hugging Face Accelerate takes a more lightweight approach: instead of filling templates automatically, it provides a step-by-step scripting interface, where users manually select torchrun parameters such as the number of processes, device mappings, and distributed backends. 

These tools make launching easier than before; as a result, the majority of large-scale DL workloads in both academia and industry today are conducted through these frameworks. But they also
have clear limits. PyTorch’s distributed launcher is usually limited to single-node execution and must rely on external launchers such as mpirun or Slurm’s srun to run across multiple nodes. In these cases, users must still write the outer launch script themselves and manually specify parameters such as node counts, ranks, communication ports, and environment variables, making the final script complex and error-prone. DeepSpeed and Accelerate do not change this mechanism but
wrap around it automatically with templates and configuration
files, they mainly support common computing environments
with limited coverage of HPC systems, and updates often
depend on the tool developers, which often prevents them from
keeping pace with new requirements

More recently, large language models (LLMs) have introduced a new paradigm for assisting users. Instead of directly managing execution like framework tools, LLMs can now assist users throughout the entire DL workloads, from setting up environments and tuning configurations to diagnosing runtime errors, lowering the barrier for non-expert users running DL jobs on different HPC systems.

Among these advances, the most relevant progress comes from code generation models and frameworks, which equip LLMs with stronger programming and reasoning capabilities directly applicable to our problem domain. These models can translate user intent into executable scripts or configuration code better than general-purpose LLMs, also including distributed launch scripts. Recent developments in this area can be broadly grouped into three main directions: model-level innovations, data-level adaptations and inference-time refinement strategies.

Model-level innovations focus on changing model architecture or fundamental generation mechanism of LLMs to better understand and generate code. For instance, StarCoder 2\cite{li2024starcoder2} introduces a Mixture-of-Experts (MoE)\cite{shazeer2017moe} architecture and a refined tokenizer to improve efficiency and cross-language understanding while scaling to larger model sizes. CRYSTAL adopts a joint encoder–decoder design that enables unified handling of both natural and programming languages, improving bidirectional reasoning between code and text \cite{zhou2024crystal}. Meanwhile, LLaDA\cite{liu2024llada} replaces the traditional autoregressive Transformer with a diffusion-based generation process, modeling code generation as a denoising task rather than sequential token prediction. This design helps the model generate more consistent and less repetitive code, particularly in long or complex sequences. 

On the other hand, data-level adaptations improve models after pre-training by fine-tuning on domain-specific data or using preference optimization. Code Llama\cite{roziere2023codellama} and CodeQwen\cite{qwen2023} are fine-tuned on large programming datasets to improve reasoning and correctness, while Magicoder\cite{ji2024magicoder} builds open-source instruction data to make generation more controllable and accurate. 

Finally, inference-time refinement strategies change how LLMs plan, generate, and verify code at runtime. Methods such as AgentCoder\cite{hu2024agentcoder} and OpenCodeInterpreter\cite{wang2024opencodeinterpreter} use multi-agent collaboration and execution feedback to iteratively test and correct code, while LDB\cite{zhang2025ldb} performs step-by-step debugging to catch and fix runtime errors. These methods enhance reliability without retraining the model.

However, these models and frameworks still lack the domain knowledge and contextual understanding required for high-performance computing environments, missing data such as parallel launch scripts, scheduler configurations, and multi-node settings. As a result, the scripts they produce may look correct but often fail to execute across real HPC systems. This gap highlights the need for new approaches that combine recent advances in code generation with HPC-specific knowledge.

\section{Conclusion}
\label{sec:conc}
Running distributed DL workloads across modern GPU clusters is a complex and error-prone task due to system heterogeneity and user-side configuration challenges. 
This work presents \name, a multi-agent system that automatically generates, verifies, and refines launch scripts for deep learning workloads. 
By combining structured template retrieval, verification, and LLM-assisted debugging, \name{} bridges the gap between user intent and system execution, reducing manual tuning and improving reproducibility.
Our experiments show that \name{} is effective across nine production GPU clusters, supports common parallelization strategies, and handles a wide range of DL models. 
It simplifies the trial-and-error procedure to fix bugs and produces structured and accurate launch scripts that can be used reliably for users with all levels of HPC expertise. 
Across the 567 test cases, \name{} achieves over 95\% accuracy, which is comparable to the state-of-the-art LLMs with template retrieval. 
\name{} is more memory efficient with 10~B paramteters, where comparable LLMs are with 100s~B parameters. 


\section{Acknowledgments}
\label{sec:acknow}
This research used resources of the National Energy Research Scientific Computing Center (NERSC), a U.S. Department of Energy Office of Science User Facility. The authors also acknowledge the Texas Advanced Computing Center (TACC) at The University of Texas at Austin for providing computational resources that contributed to the results presented in this paper. URL: https://www.tacc.utexas.edu.

This work additionally utilized Delta and DeltaAI resources at the National Center for Supercomputing Applications (NCSA) through support from the ACCESS program. The authors gratefully acknowledge the use of Delta (https://delta.ncsa.illinois.edu/delta-citations/
) and DeltaAI (https://delta.ncsa.illinois.edu/delta-ai-citation/
) systems, as well as support provided via ACCESS, NAIRR Pilot, and Illinois Computes.

This research is supported by the NSF OAC-2401245 and OAC-2411294.

\balance
\bibliographystyle{plain}
\bibliography{MPEA}

\end{document}